\documentclass[twocolumn,trackchanges]{aastex7}

\usepackage{graphicx} 

\usepackage{enumitem}
\usepackage{orcidlink}
\usepackage{multirow}
\usepackage{amsmath}
\usepackage{upgreek}

\usepackage{placeins}
\usepackage{float}

\usepackage{chngpage}
\usepackage{tikz}
\usetikzlibrary{arrows,calc,positioning}
\tikzstyle{intt}=[draw,text centered,minimum size=6em,text width=5.25cm,text height=0.34cm]
\tikzstyle{intl}=[draw,text centered,minimum size=2em,text width=2.75cm,text height=0.34cm]
\tikzstyle{int}=[draw,minimum size=2.5em,text centered,text width=3.5cm]
\tikzstyle{intg}=[draw,minimum size=3em,text centered,text width=6.cm]
\tikzstyle{sum}=[draw,shape=circle,inner sep=2pt,text centered,node distance=3.5cm]
\tikzstyle{summ}=[drawshape=circle,inner sep=4pt,text centered,node distance=3.cm]
\usetikzlibrary{shapes.geometric, arrows}
\tikzstyle{arrow} = [thick,->,>=stealth]

\shorttitle{Auroral Thermal Inversions}
\shortauthors{Zuckerman \& Mang et al.}

\graphicspath{{./}{figures/}}

\begin{document}

\title{Can Auroral Heating Explain Brown Dwarf Thermal Inversions?}

\correspondingauthor{Zuckerman}
\email{anna\_zuckerman@alumni.brown.edu}

\newcommand{\CUB}{Department of Astrophysical and Planetary Sciences, University of Colorado Boulder, 2000 Colorado Ave, Boulder, CO 80309}
\newcommand{\UT}{Department of Astronomy, University of Texas at Austin, 2515 Speedway, Austin, TX 78712}
\newcommand{\LASP}{Laboratory for Atmospheric and Space Physics, 3665 Discovery Drive Boulder, CO 80303}

\author[0000-0002-2412-517X]{Anna Zuckerman}
\altaffiliation{These authors contributed equally to this work}
\altaffiliation{NSF Graduate Research Fellow}
\affiliation{\CUB}
\affiliation{\LASP}
\email{}

\author[0000-0001-5864-9599]{James Mang}
\altaffiliation{These authors contributed equally to this work}
\altaffiliation{NSF Graduate Research Fellow}
\affiliation{\UT}
\email{}

\author[0000-0002-4489-0135]{J. Sebastian Pineda}
\affiliation{\CUB}
\affiliation{\LASP}
\email{}

\author[0000-0002-4404-0456]{Caroline V. Morley}
\affiliation{\UT}
\email{}

\author[0000-0001-8932-368X]{David Brain}
\affiliation{\CUB}
\affiliation{\LASP}
\email{}

\begin{abstract}
Recent modeling of the JWST spectra of the Y-dwarf CWISEPJ193518.59-154620.3 \citep{Faherty2024, Suarez2025} and the T-dwarf SIMPJ013656.5+093347.3 \citep{Nasedkin2025} indicates the presence of a thermal inversion in their atmospheres, implying an additional source of heating. Auroral energy deposition has been proposed as a potential mechanism. In this work, we model auroral energy deposition processes and the resulting atmospheric response to assess whether physically motivated auroral electron beam energy spectra can reproduce the inferred thermal inversions. We simulate the effects of several auroral electron beam energy spectra consistent with observations of the Jovian population of auroral electrons and scale them to the total electron flux expected for brown dwarfs. We find that auroral energy deposition alone cannot reproduce both the magnitude and pressure level of the inferred inversion in either object. While inversions are produced in our models, they occur at pressures that are too low compared to those required by the observations. These results suggest that the actual population of auroral electrons includes a stronger high-energy component than considered here, that additional heating mechanisms contribute to the observed thermal inversions, and/or that heterogeneous heating of the atmosphere is important for determining the observed emission.
\end{abstract}


\section{Introduction}
\label{sec:motivation}


Brown dwarfs offer rich laboratories for studying the interactions of atmospheres and magnetospheric phenomena. Increasingly powerful observational capabilities have opened questions which require new modeling and theoretical work to explain. In particular, recent observations indicate thermal inversions in the atmospheres of some brown dwarfs, which require an unexplained additional source of external heating. In this work we combine atmospheric and auroral modeling to ask whether these inversions could be explained by auroral processes.

\cite{Faherty2024} presented JWST observations of an unexpected methane feature in the spectrum of CWISEPJ193518.59-154620.3 (abbreviated to WISE1935 in this paper), a 482$\pm38$~K free-floating Y-type brown dwarf. The JWST spectrum of WISE1935 displayed an emission feature in the CH$_4$ Q-branch centered at 3.326 $\upmu$m. The Y-dwarf WISEJ222055.31-362817.4, observed in the same program and sharing near identical fundamental parameters (ie. luminosity, effective temperature, surface gravity) to WISE1935, showed only CH$_4$ absorption. \cite{Faherty2024} used the \texttt{Brewster} atmospheric retrieval framework \citep{Burningham2017, Burningham2021} to model this emission feature. The pressure-temperature (P-T) profile that best fits the spectrum for WISE1935 shows an approximately 300K temperature inversion centered at 1-10 mbar. The authors also tested a retrieval without allowing for the presence of a thermal inversion, and found that it could not reproduce the methane emission feature. Thus, \cite{Faherty2024} concluded that a thermal inversion is present in the atmosphere of WISE1935, despite no significant external source of stellar irradiation to provide a source of heating. Though WISE1935 has been recently shown to have a substellar companion, its low temperature ($\sim$350K) is insufficient to produce significant heating \citep{DeFurio2025}. \cite{Faherty2024} mentions that this may be a similar phenomenon to the unexpectedly hot upper atmospheres of solar system gas giants. In gas giants, several possible mechanisms for heating of the upper atmosphere have been hypothesized, including auroral energy deposition and transport of energy from deeper in the atmosphere by vertically propagating waves. If energy transport from lower layers were the primary mechanism, they argue that thermal inversions should be common among such Y-dwarfs, and so the fact that inversions are not commonly observed suggests auroral processes may be important. Auroral energy deposition at polar latitudes could be redistributed across the object, as occurs on Jupiter \citep{O'Donoghue2021}. 

A similar inversion has been reported in SIMPJ013656.5+093347.3 (abbreviated SIMP0136 in this paper) \citep{Nasedkin2025}, a $\sim1245$~K, isolated T-dwarf. Atmospheric retrievals conducted with \texttt{petitRADTRANS} \citep{Molliere2019} using JWST/NIRSpec and MIRI/LRS spectra of this object showed a $\sim$265 K inversion in the retrieved temperature profile peaking near 3 mbar. This retrieved inversion was found to be highly sensitive to the observed methane absorption features at 3.3 and 7.7 $\upmu$m. \cite{Nasedkin2025} argue that if the extrapolation of Jovian auroral energy deposition presented by \cite{Pineda2024} holds, auroral precipitation is a plausible heating mechanism for this inversion. They then make an energy-balance based argument that the expected auroral energy deposition on its own may not be sufficient to directly produce the observed excess emitted power unless the auroral energy deposition mechanisms differ from the Jovian analog. Their simple approach supports the need for the more detailed modeling presented here. 

Several lines of evidence confirm that auroral energy deposition is common in the atmospheres of brown dwarfs. Circularly polarized pulsed radio emission has been observed may times from individual brown dwarfs \citep[e.g.,][]{berger2002, Burgasser2005, Berger2005, Berger2006, Hallinan2006, Hallinan2007, Route2012, Williams2015, Kao2018, Vedantham2020, Vedantham2023, Rose2023, Guirado2025, Miles-Paez2025,  Wandia2026}, and has been shown to be consistent with the Electron-Cyclotron Maser Instability (ECMI) emission expected from aurorally precipitating energetic electrons \citep{Hallinan2008}. Subsequent observations indicate brown dwarfs with H$\alpha$ emission are more likely to have such radio detections \citep{Kao2016}, supporting the idea that the auroral electron beam interacts with the atmosphere upon reaching it as established by \cite{Hallinan_2015}. Ongoing observational work \citep[e.g.][]{Gibbs2022, Pineda2024} is searching for the additional infrared and ultraviolet auroral emission expected from this auroral precipitation \citep{Pineda2017}.

These energetic precipitating auroral electrons lose their energy through interactions with atmospheric molecules. In H$_2$ dominated brown dwarf atmospheres, much of their energy is lost to electronic excitations and ionizations of H$_2$ which are expected to contribute to observable auroral emission. Each electron will eventually lose sufficient energy such that ionizations and excitations are no longer possible. The energy remaining contributes to heating, as does energy lost to vibrational and rotational excitations of H$_2$. This heating might then influence the thermal structure of the atmosphere. 

Solar system gas giants provide our closest observational analogs for understanding auroral processes on brown dwarfs. In particular, Jupiter has an inversion in its upper atmosphere which has also been proposed to be due at least in part to auroral heating \citep{O'Donoghue2021}. Because of the relative observational accessibility of Jupiter compared with WISE1935 and SIMP0136, we can leverage in situ Jovian observations as an initial framework for understanding auroral processes on these objects.

In this work, we seek to determine whether auroral energy deposition can provide the heating required to produce the thermal inversions implied for WISE1935 and SIMP0136. In Section \ref{subsec:picaso}, we describe the atmospheric model used to define the initial atmospheric profiles as well as to simulate the atmospheric response to heating. In Section \ref{subsec:energy_dep}, we describe the model for energy deposition in brown dwarf atmospheres. In Section~\ref{subsec:beam_spectrum}, we derive the expected auroral electron beam energy spectrum incident on atmospheres like those of WISE1935 and SIMP0136. In Section \ref{subsec:energy_flux} we describe the calculation of the energy flux through the atmosphere resulting from these auroral beams. In Section \ref{subsec:atmospheric_modeling}, we present our framework for simulating the atmospheric response to this energy flux. The results of these models are presented in Section~\ref{sec:results}. In Section~\ref{sec:discussion}, we examine the implications of our findings, including model assumptions and alternative heating mechanisms. We summarize our findings in Section \ref{sec:conclusion}.
 
\section{Methods}

We model the heating resulting from the interaction of an auroral electron beam with brown dwarf atmospheres using the parameterized auroral model of \cite{Zuckerman2026}. We then calculate the atmospheric response to this heating using custom \texttt{PICASO} models. The details of each component are described in the following sections.

\subsection{\texttt{PICASO} \label{subsec:picaso}}

In this work, we generate auroral models and the response of the atmosphere to these auroral profiles using custom atmospheric models generated with \texttt{PICASO} \citep{Mukherjee2023, Mang2026}. \texttt{PICASO} is an open-source package that can generate 1D radiative–convective equilibrium (RCE) models and simulate spectra for brown dwarfs and exoplanets in chemical equilibrium and disequilibrium. \texttt{PICASO} has been used in numerous studies to model brown dwarfs and exoplanets \citep[e.g.][]{Miles2023, Powell2024, Crotts2025, Matthews2026, Sanghi2026, Smith2026}. 

In this work, we assume all atmospheres are in chemical disequilibrium, as recent studies of substellar objects show strong evidence of this \citep{Miles2020, Zhang2025, Nasedkin2025}. The strength of vertical mixing in these models is parameterized using the eddy diffusion coefficient, $K_{\rm zz}$. All of the models presented here are generated with $K_{\rm zz} = 10^7$ cm$^2$ s$^{-1}$. The implications of assuming chemical disequilibrium and the choice of $K_{\rm zz}$ are further discussed later (Section \ref{sec:deq-chem}). 

For our WISE1935 models, we use an effective temperature of $T_{\rm eff} = 482$~K, surface gravity of log(g) = 4.7 (cgs), and metallicity of [M/H] = +0.5, the median values from \citet{Faherty2024}. Note that in our models, we follow the C/O ratio from \citet{Lodders2020} where solar C/O = 0.549. For our models of SIMP0136, we set the model parameters to be $T_{\rm eff} = 1245$~K, log(g) = 4.5, [M/H] = +0.0, and C/O = 0.549.

We also use \texttt{PICASO} to compute simulated thermal emission spectra. Opacities are taken from \citet{Freedman2008}, with updates from \citet{Freedman2014}, with further detail provided in \citet{Mang2026}. These spectra are generated at a resolving power of $R = 50{,}000$, sufficiently high to allow accurate binning and convolution to the JWST observations of WISE1935 (NIRSpec/G395M) and SIMP0136 (NIRSpec/PRISM). For each object, we convolve the model spectra to the instrument's resolving power and rebin onto the JWST data wavelength grid.

Since \texttt{PICASO} is a 1D climate model, we cannot directly model higher-dimensional effects such as horizontal mixing, although we approximate its effect in Section \ref{sec:spatial}. Similarly, our models cannot fully capture all modes of vertical energy transport, for instance vertical downwelling over the poles. We discuss these limitations further in Section \ref{subsec:1Dvs3D}.

\subsection{Auroral Model}
\label{subsec:energy_dep}

We use a recently developed parameterization of the energy deposition profile resulting from the precipitation of auroral electrons into substellar atmospheres \citep{Zuckerman2026} to model auroral energy deposition in the two brown dwarf atmospheres. This parameterization was derived from Monte Carlo simulations of energetic electrons interacting with substellar atmospheres. It assumes electrons interact primarily with H$_2$ and requires as input an assumed atmospheric density profile and electron beam energy spectrum. 

We derive density profiles and a relationship between altitude and pressure by interpolating the temperatures, pressures, and mixing ratios produced by the \texttt{PICASO} model, then numerically integrating assuming an ideal gas and defining the zero-point in altitude at a pressure of one bar. The \texttt{PICASO} model is valid below the homopause, approximately 10$^{-6}$ bars. At lower pressures, we extend the atmosphere isothermally. We take the density profile in this isothermal region to decay from the value at the homopause as governed by hydrostatic equilibrium.

\subsection{Electron Beam Energy Spectrum}
\label{subsec:beam_spectrum}
The second input required to calculate the parameterized energy deposition rate is the expected electron beam energy spectrum. Although the mechanism producing the auroral electron beams for WISE1935 and SIMP0136 are unknown, the beam spectrum can be approximated based on Jovian measurements and estimates of the total auroral flux based on radio observations.

In order to model physically reasonable auroral energy deposition rates, we need to understand the expected auroral electron beam energy spectrum for brown dwarfs. Our closest solar system analog, Jupiter, is a useful starting point because similar processes may underlie the production of the auroral electron beam in the brown dwarf and Jovian cases \citep{Hallinan_2015}. We emphasize that, as the best analogy for which in situ measurements are possible, the Jovian system provides an informative reference point, but of course the measured Jovian electron beam will differ from the actual beam in the brown dwarf case. The most recent observations come from the Juno mission, which has measured electron populations at various magnetic latitudes, altitudes, and pitch angles across its many orbits of Jupiter \citep[e.g.,][]{Mauk2017b, Mauk2018, mauk2020}. \cite{Mauk2024} used measurements of the energetic electron population detected by Juno's Jupiter Energetic Particle Detector Investigation (JEDI) and Magnetometer Investigation's Advanced Stellar Compass (ASC) instruments to model the energy spectrum of the electrons in these intense, bi-directional auroral electron beams found at high Jovian latitudes. Though the electron population differs across time, latitude within the auroral oval, and upwards or downwards direction of travel, its spectrum is in general fit by a kappa distribution plus power law tail in each observation. \cite{Mauk2024} used the following form to fit this distribution:

\begin{equation}
\label{eq:Mauk}
    F = C \varepsilon_0 \frac{[kT(g1 + 1) + \varepsilon_0)]^{-g1 - 1}}{1+(\varepsilon_0/\varepsilon_1)^{g2}}
\end{equation}

\begin{figure}[ht]
\centering
\includegraphics[width=\columnwidth]{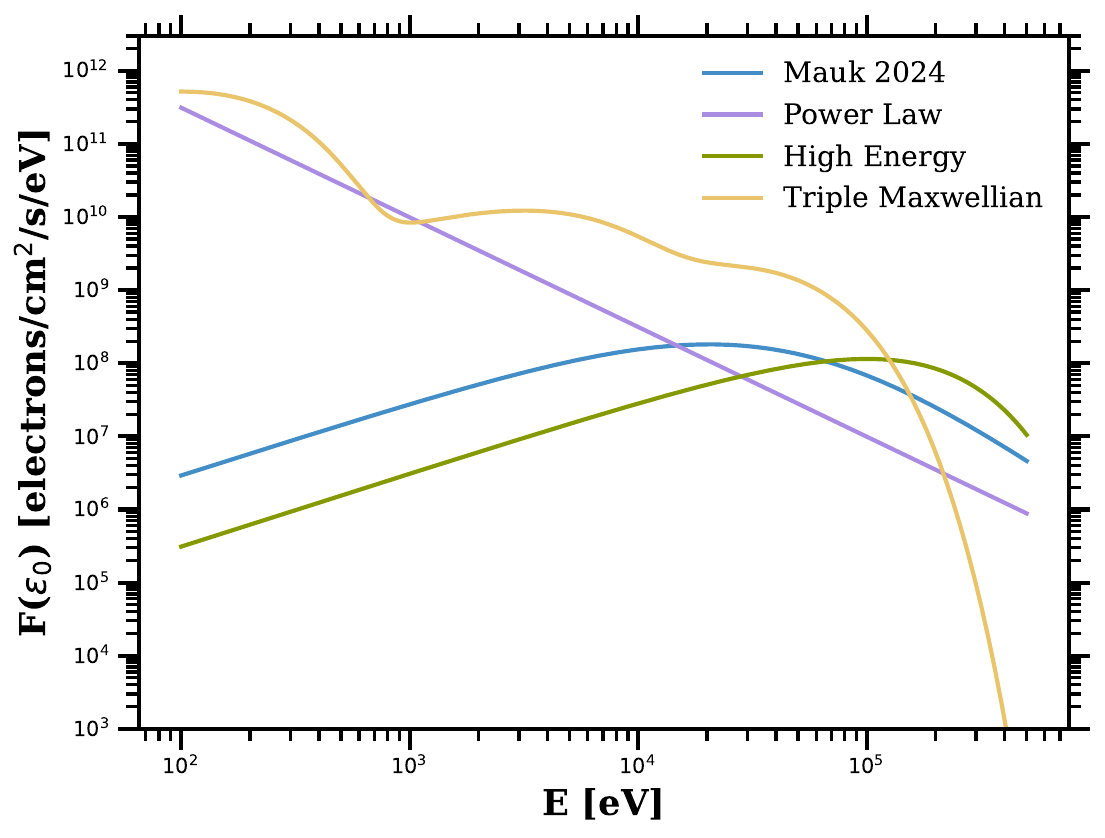}
\caption{The four representative electron beam energy spectra used in this work, scaled to produce a total incident energy flux consistent with observed brown dwarf ECMI radio emission. The spectrum labeled Mauk 2024 has the form presented in Eq.\ \ref{eq:Mauk} with parameters $k=14.1$, $g1=2.26$, $g2=0$, and $\varepsilon_0=1$ taken from \cite{Mauk2024}, and $C=1.55e16$ cm$^{-2}$ s$^{-1}$ keV$^{-1}$ from scaling the Jovian value of $C$ as described in Section \ref{subsec:beam_spectrum}. The power law spectrum is based on underlying Alfv\'en wave physics, the Triple-Maxwellian spectrum is motivated by pre-Juno Galileo observations, and the high-energy spectrum is designed to test the effects of a strong high-energy beam component.}
\label{fig:beam_spectra}
\end{figure}

\noindent This expression gives the electron flux $F$ in electrons cm$^{-2}$ s$^{-1}$ keV$^{-1}$. The energy terms $kT$, $\varepsilon_0$, and $\varepsilon_1$ are expressed per keV and so made dimensionless, $g1$ and $g2$ are dimensionless fitting parameters, and $C$ has units of cm$^{-2}$ s$^{-1}$ keV$^{-1}$. \cite{Mauk2024} fit these parameters for each observation they discuss. For a specific representative example, we took the parameters from their observation 133329-L because it represents typical count rates in the ASC detector and was taken when the spacecraft was probing high-latitude regions in the magnetic field of Jupiter inside the auroral oval, but we note that the numerical value will be subsequently scaled regardless of the initially chosen parameters. This spectrum is shown in Figure~\ref{fig:beam_spectra}, along with several others as described later in this section.

Next, we estimated the total electron flux expected to be carried by brown dwarf auroral electron beams. Observations of auroral radio emission from brown dwarfs have been used to constrain the overall auroral electron flux. For instance, \cite{Turnpenney2017} modeled the precipitating auroral electron flux under two scenarios: a Jovian picture in which plasma generated by a satellite flows outwards in a closed magnetosphere, and a model in which a rotating open magnetosphere interacts with the interstellar medium. They found that the open magnetosphere model can match observed radio luminosities across a broader range of parameter space. \cite{Turnpenney2017} present plots of the calculated precipitating electron energy flux as a function of latitude. Taking the energy flux as a constant value over the narrow auroral region provides an upper limit to the total expected flux. In the open magnetosphere model, the electron energy flux consistent with observed ECMI radio emission reaches the beam's maximum at about 10 kW m$^{-2}$. We can then enforce that the total flux resulting from each of our electron energy spectra matches this value. Integrating the spectrum in Equation \ref{eq:Mauk} weighted by energy over the approximate expected 1e-1 to 1e5~keV range gives the energy flux expected from Jupiter, which allows us to scale that spectrum to produce a flux consistent with observed radio emission as calculated by \cite{Turnpenney2017}. We found that scaling the \cite{Mauk2024} spectrum upwards by just over six orders of magnitude approximates the fluxes calculated by \cite{Turnpenney2017}. We use this scaled spectrum based on \cite{Mauk2024} measurements as our baseline energy spectrum.

The actual population of auroral electrons for brown dwarfs is not well constrained. We can make a rough estimate based on the above Jovian analogy, but that will inherently not represent the real picture for any specific object. For that reason, we tested a series of forms for the energy spectra based on existing literature to assess the impact of various auroral electron populations. In addition to the Juno-informed baseline spectrum described above, we also tested three additional forms used in earlier modeling works: (1) a power-law spectrum grounded in the expected underlying physics of Alfv\'en wave turbulence \citep{Saur2002}, (2) a Triple-Maxwellian spectrum motivated by pre-Juno Galileo Extreme Ultraviolet Spectrometer observations \citep{Grodent2001}, and (3) a high-energy spectrum designed to test the effects of a strong high-energy beam component. The beam energy spectra tested are shown in Figure \ref{fig:beam_spectra}.

\subsection{Energy Flux Rate}
\label{subsec:energy_flux}

Using these beam energy spectra and the modeled atmospheric density profiles, the parameterization derived by \cite{Zuckerman2026} provides an expression for the total volumetric event rates for each interaction type, $Q_{event}$ [events cm$^{-3}$ s$^{-1}$], and for energy deposition, $Q_{\varepsilon}$ [eV cm$^{-3}$ s$^{-1}$]:

\begin{equation}
\label{eq:Qenergy}
    Q_{\varepsilon}(z) = \int q_{\varepsilon}(\varepsilon_0, z) F(\varepsilon_0) d\varepsilon_0 \; ,
\end{equation}

\noindent where $\varepsilon_0$ is the incident electron energy, $z$ represents altitude within the atmosphere (here defined such that $P(z=0) = 1$ bar) and $F(\varepsilon_0)$ is the energy spectrum of the incident beam (in electrons cm$^{-2}$ s$^{-1}$ eV$^{-1}$). $q_{\varepsilon}$ is given by

\begin{equation}
\label{eq:qenergy}
    q_{\varepsilon} (z) = C_{\varepsilon} \beta_{ion}(\varepsilon_0) \mathcal{P}(N_{ion}(z)|\varepsilon_{0})n(z)
\end{equation}

\noindent where $\beta_{ion}$  is the ratio of the total number of ionizations to the number of electrons in the incident beam as a function of energy and $n(z)$ is the number density of H$_2$. Values for $C_{\varepsilon}$ and $\beta_{ion}$ are given in \cite{Zuckerman2026}. $N_{ion}$ is the column density at which ionizations occur (the integral of the number density profile over altitude), and $\mathcal{P}(N_{\varepsilon}(z)|\varepsilon_{0})$ is the probability distribution for $N_{ion}$,

\begin{equation}
\label{eq:pdf}
    \mathcal{P}(N_{ion}) = \frac{1}{\sigma\sqrt{2\pi}}N_{ion}^{\frac{1}{2\sigma} - 1}\exp\left[\frac{\mu }{2\sigma } - \frac{1}{2}N_{ion}^{1/\sigma} e^{\frac{\mu }{\sigma }}\right]
\end{equation}

\noindent where $\mu$ and $\sigma$ are given in \citet{Zuckerman2026}.

\begin{figure*}[ht]
\centering
\includegraphics[width=\textwidth]{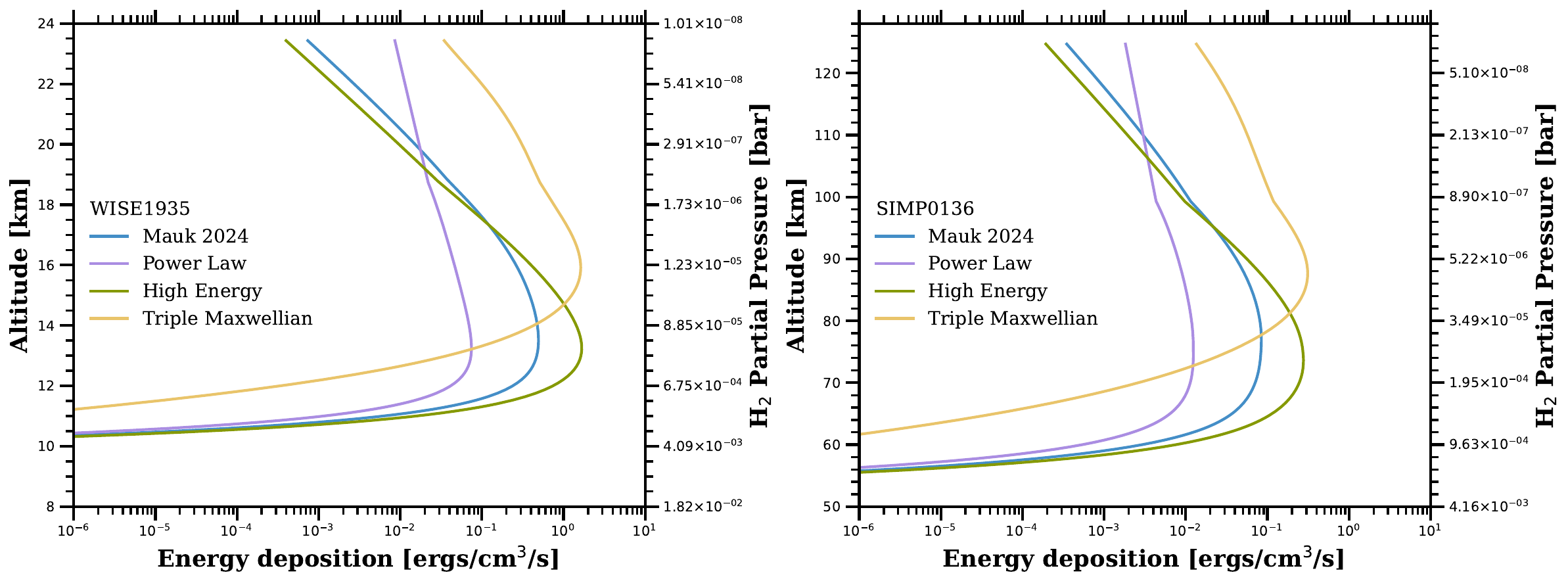}
\caption{Energy deposition rates resulting from each of the tested beam spectra for WISE1935 (left panel) and SIMP0136 (right panel). The shapes are very similar in pressure space but span different altitudes predominantly due to the higher gravity of WISE1935 compared to SIMP0136}
\label{fig:energy_deposition}
\end{figure*}

The energy deposition rate $Q_{\varepsilon}$ [ergs cm$^{-3}$ s$^{-1}$] is shown in Figure \ref{fig:energy_deposition}. Using this parametrization of $Q_{\varepsilon}$, we can then calculate the energy flux through a given layer of the atmosphere. As discussed in Section \ref{subsec:atmospheric_modeling}, this is necessary as an input for modeling of the atmospheric response to the energy deposition. The energy flux $E_{\varepsilon}$ [ergs cm$^{-2}$ s$^{-1}$] is calculated as the integral of $Q_{\varepsilon}$ over each layer of the atmosphere. 

Because the actual mechanism for generating the auroral electron beam is unknown for WISE1935 and SIMP0136, we treated the numerical value of $E_{\varepsilon}$ [ergs cm$^{-2}$ s$^{-1}$] as an approximation supported by the currently available literature and varied it by several orders of magnitude to assess the affect of different beam intensities for different beam spectra. 
    
\subsection{Atmospheric Response}
\label{subsec:atmospheric_modeling}

The atmosphere responds to the additional energy deposited by auroral precipitation through a combination of chemical reactions and energy transport processes, which alter the atmospheric composition and redistribute energy, ultimately modifying the pressure--temperature profile. We estimate the impact of these processes by injecting energy (as calculated in Section \ref{subsec:energy_flux}) into layers of the atmosphere, as we calculate its structure. To model the response of the atmosphere, we use the \texttt{PICASO} energy injection module, which allows an external energy flux to be deposited into the atmosphere. There are two ways within \texttt{PICASO} to do this. The first is using a user-defined energy deposition profile. We use this option to directly inject the electron beam energy profiles, described in the previous section, and allow the models to reach radiative-convective equilibrium to capture the resulting thermal structure.

The second method in \texttt{PICASO} to inject energy is using an ad hoc energy deposition profile in the form of a Chapman function. This ad hoc function, a parameterized vertical heating profile, has been used to represent atmospheric heating from sources of opacity not explicitly included in the model. In practice, it produces a peaked heating profile in pressure space centered at a specified pressure level, with a vertical extent set by a characteristic scale height. The resulting profile resembles a smooth, approximately Gaussian-like distribution in log-pressure, although it is derived from an exponential attenuation of incident flux with optical depth. This approach has been adopted in previous studies \citep{MarleyMcKay1999, Morley2014patchy}. For all Chapman profiles considered here, we assume the energy deposition extends over one atmospheric scale height. So in addition to the auroral energy profiles, we also employ the Chapman function to investigate the range of energy required to recreate the observed thermal inversions in WISE1935 and SIMP0136.

\section{Results}
\label{sec:results}

\subsection{WISE 1935}
\label{subsec:WISE1935}

\subsubsection{Response to Physically-motivated Electron Energy Spectra}
\label{subsec:WISE1935_baseline_spectra}

\begin{figure*}[ht!]
\centering
\includegraphics[width=\textwidth]{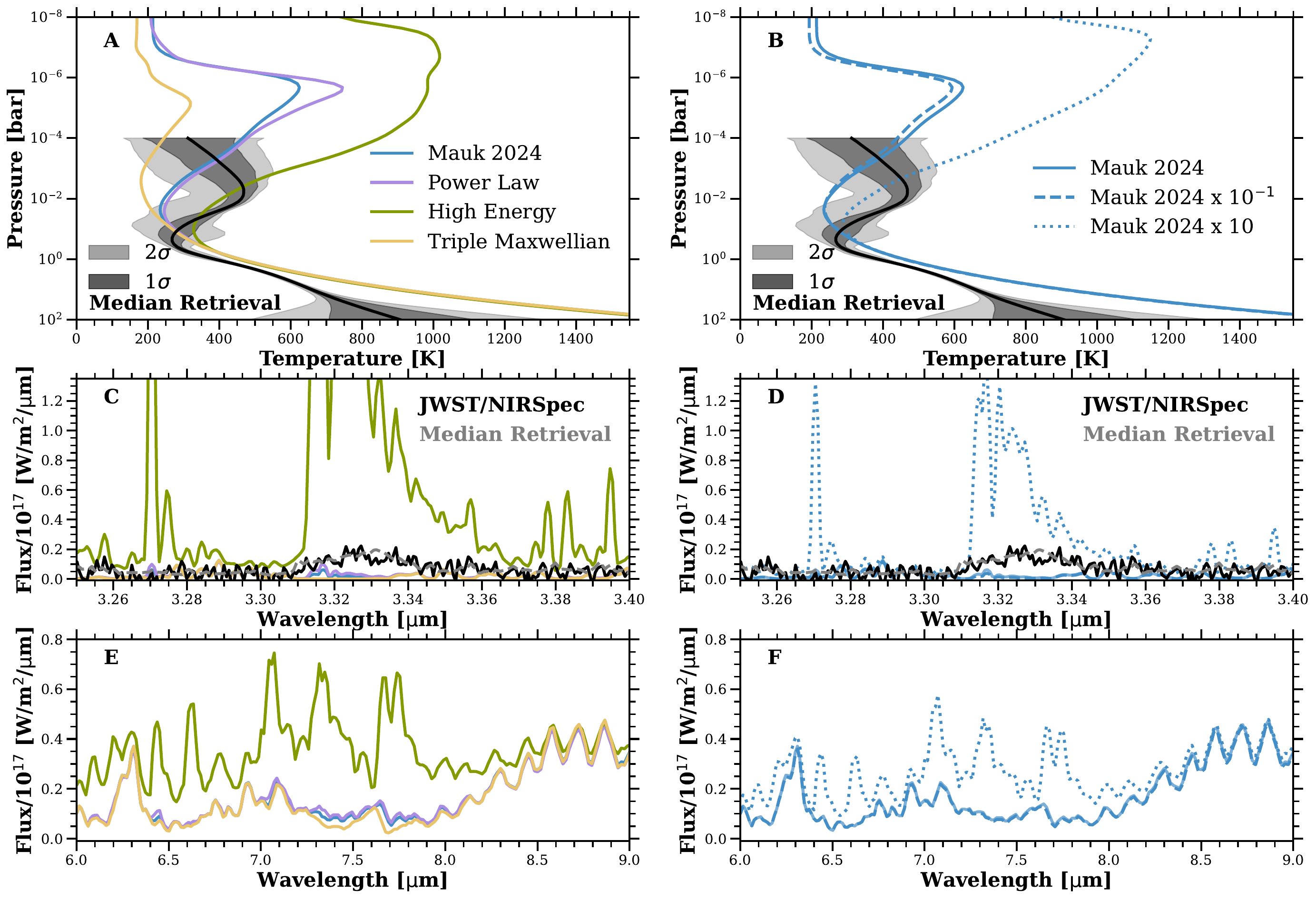}
\caption{A: Pressure–temperature profiles of WISE1935. The median retrieved profile from \citet{Faherty2024} with a thermal inversion is shown in black, with the 1$\sigma$ and 2$\sigma$ regions shaded in gray. Solid colored lines show \texttt{PICASO} models with different energy deposition profiles. 
B: Same as panel A, but for models using the baseline beam energy spectrum scaled down by a factor of 10 (dashed) and scaled up by a factor of 10 (dotted). 
C: CH$_4$ spectral region near $\sim$3.33~$\upmu$m observed with JWST/NIRSpec (black), compared to model spectra including auroral energy deposition from each beam spectrum and the median retrieved spectrum (gray). 
D: Same as panel C, but for models with the baseline beam spectrum scaled by factors of 10. 
E: CH$_4$ spectral region from 6–9~$\upmu$m for models with different auroral energy deposition profiles. 
F: Same as panel E, but for models using the scaled Mauk spectrum.}

\label{fig:WISE1935pt}
\end{figure*}

Figure \ref{fig:WISE1935pt} shows the thermal structure profiles and spectra corresponding to the models that we generate with the different electron beam energy deposition profiles. We compare these to the retrieved profile with a thermal inversion of WISE1935 in \citet{Faherty2024}. The black profile represents the median retrieved P-T profile with a thermal inversion, and the gray shaded areas represent the 1 and 2$\sigma$ regions. We find that no electron beam energy spectrum fully reproduces the observed thermal inversion in WISE1935 (Figure~\ref{fig:WISE1935pt}, panel A). Although all beam models generate inversions, these occur at significantly lower pressures than the retrieved profile. The high-energy beam produces the closest match in inversion strength, but still fails to reproduce the location of the inversion near $\sim 10^{-2}$ bar and continues to exceed the inversion strength at much lower pressures. Correspondingly, none of the beam models reproduce the observed CH$_4$ emission feature near $\sim$ 3.30 to 3.36 $\upmu$m in the JWST/NIRSpec data in black (Figure~\ref{fig:WISE1935pt}, panel C).

We further test whether varying the total deposited energy can reconcile these differences by scaling the baseline \citet{Mauk2024} energy deposition profile (Figure~\ref{fig:WISE1935pt}, panel B). While increasing the total energy increases the strength of the thermal inversion, it does not alter the vertical extent or altitude of the inversion. Even when scaled by a factor of 10, the resulting profiles remain inconsistent with the retrieved inversion structure near $10^{-2}$ bar, similar to the fiducial high energy deposition model. As shown in panel D, the corresponding spectra also fail to reproduce the observed CH$_4$ emission feature.

Additionally, the high-energy electron beam model produces emission features near 3.27 and 3.39~$\upmu$m that are not present in the observed spectrum. If these beam profiles were responsible for the inversion, such features would be expected within the CH$_4$ bandpass, further indicating the discrepancy between the electron-deposited heating and the observed spectra. The source of these additional features is discussed in Section \ref{sec:methane_lineshape}.

\subsubsection{Response to Ad-Hoc Heating Profile}
\label{subsec:WISE1935_ad_hoc_spectra}

If we set aside the need for a physically motivated auroral mechanism and instead allow both the magnitude and vertical location of the heating to vary using the ad hoc energy deposition described in Section \ref{subsec:atmospheric_modeling}, we can reproduce the retrieved thermal inversion of WISE1935 reasonably well, as seen in Figure \ref{fig:WISE1935chapman}. A profile depositing $\sim 4 \times 10^{5}$ erg~cm$^{-2}$~s$^{-1}$ ($\sim 13\%$ of the bolometric flux of WISE1935) with a peak at 0.01 bar reproduces both the amplitude and approximate pressure location of the retrieved inversion. This demonstrates that the primary limitation of the electron beam models is not simply the total amount of deposited energy, but the pressure level at which that energy is deposited. Since the penetration depth of auroral electrons is controlled by the electron energy distribution, reproducing the observed inversion with auroral precipitation would require a substantially stronger high-energy component than is present in the fiducial beam models. The presence of such high-energy tails is in theory consistent with both Jovian observations and theoretical predictions (Section~\ref{sec:highenergy}).

The ad hoc energy deposition models also more closely reproduce the observed CH$_4$ emission feature. In particular, they match both the amplitude and the broader spectral shape across the 3.31--3.35~$\upmu$m region. In contrast, the electron beam models produce a narrower feature that declines near 3.33~$\upmu$m, failing to capture the observed spectral morphology. We defer further discussion of this discrepancy to Section~\ref{sec:methane_lineshape}.

\subsubsection{Beam Modification to Reproduce Required Heating}
\label{subsec:WISE1935_mods}
As discussed above, we are able to simulate the inferred inversions with heating taking the form of an ad hoc Chapman heating profile. We have shown that none of the physically motivated electron beam energy spectra are able to reproduce this heating on their own. We can attempt to modify the energy spectrum to achieve the required heating profile.

Because the heating from our electron beams occurs at lower pressures than the necessary heating for the inversions, and because higher energy electrons deposit their energy at higher pressures, we experimented with both extending the baseline Juno-informed energy spectrum to higher energies and with shifting the peak of that spectrum to higher energies. We emphasize that any simulation of electrons above about 500 keV is approximate because we do not treat electrons relativistically, as discussed above. The modified beams discussed in this section and in Section \ref{subsec:SIMP0136_mods} were simulated by simply extrapolating the non-relativistic parameterization to higher energies, as discussed in Appendix \ref{sec:extrapolation}.

Neither extending nor shifting the beam energy spectrum can reproduce the inversions, because they produce heating profiles that are too wide. This is due primarily to the electrons in the low-energy tail of the energy distribution, which deposit their energy across a much wider range of low pressures than required for the inversion even when the spectrum is extended or shifted so that peak energy deposition occurs at the correct pressure level.

\begin{figure}[ht!]
\centering
\includegraphics[width=\columnwidth]{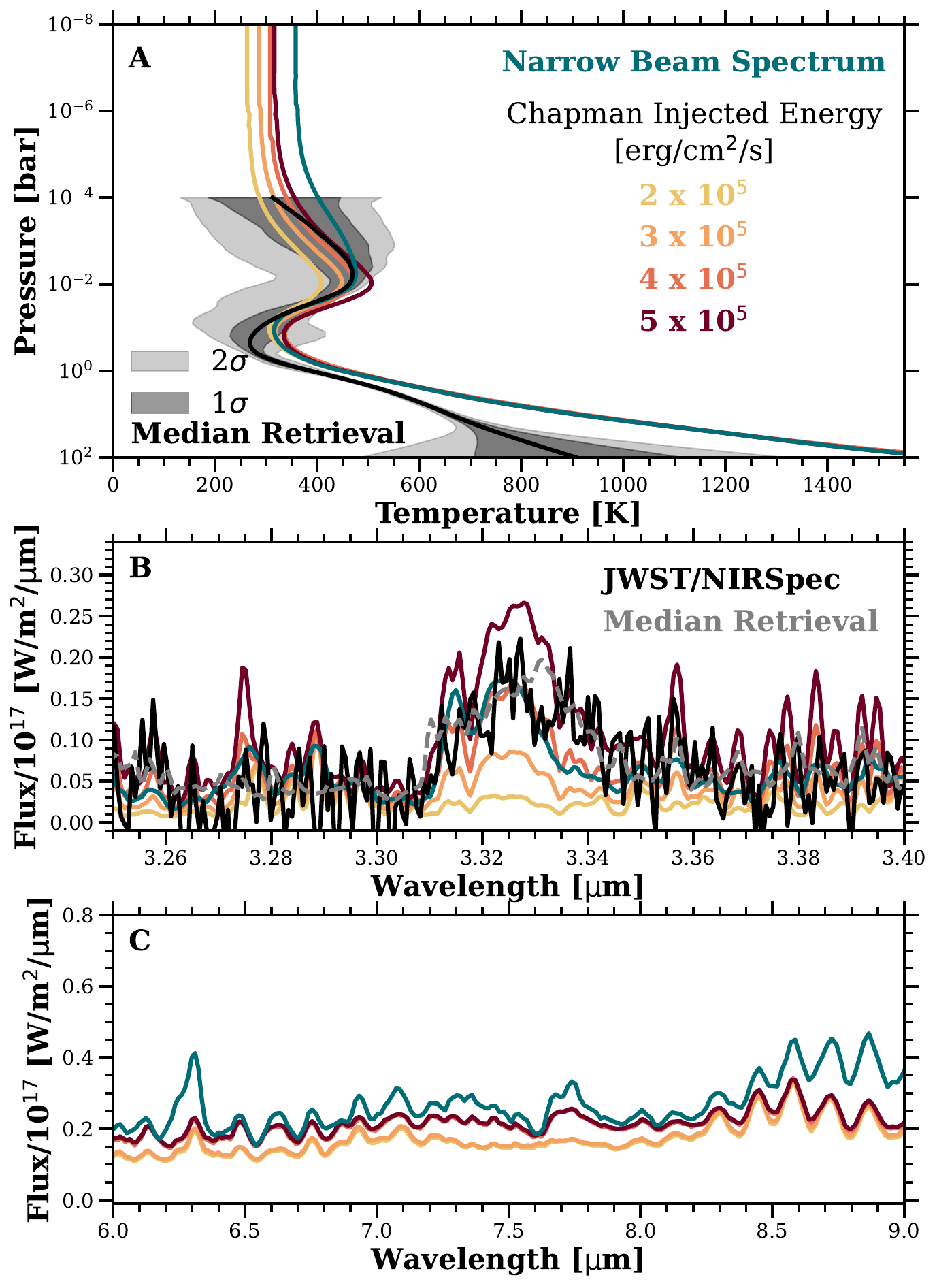}
\caption{A: Simulated pressure–temperature profiles of WISE1935 with energy injected using an ad hoc heating profile peaking at 0.01 bar. Each profile is labeled by the total injected energy flux (erg cm$^{-2}$ s$^{-1}$). The profile with a modified, narrow electron beam spectrum is shown in blue.
B: CH$_4$ spectral region near $\sim$3.33~$\upmu$m observed with JWST/NIRSpec (black), compared to models with Chapman function energy injection at varying total fluxes and the median retrieved spectrum (gray). 
C: CH$_4$ spectral region from 6–9~$\upmu$m for the same set of models, corresponding to wavelengths in future JWST observations.}
\label{fig:WISE1935chapman}
\end{figure}

We are able to achieve a heating profile close to the required ad hoc profile if we take a very narrow, high-energy electron distribution. The best profile was produced with a narrow slice of the baseline spectrum shifted such that the peak of the spectrum occurs at 1500keV, spanning 0.95 to 1.05 the value of the peak in range, and scaled down by an additional 40\% from the baseline spectrum (defined in Section \ref{subsec:beam_spectrum}). This energy spectrum produces a heating profile which peaks at the correct pressure but is somewhat wider than the ad hoc heating profile, especially on the low-pressure end (as shown in Appendix \ref{sec:modified_beam_heating}). As shown in Figure \ref{fig:WISE1935chapman}, the resulting temperature profile from this energy spectrum has an inversion at approximately the required location, but results in overall higher temperatures at low pressures.

This energy spectrum is unphysical as there is no clear mechanism for producing such a narrow, high-energy spike in the spectrum. This underscores that the simple picture of heating from auroral energy deposition from the population of electrons we have modeled is insufficient to produce the heating required for the inferred inversions on its own.

\subsection{SIMP 0136}
\label{subsec:SIMP0136}

\subsubsection{Response to Physically-motivated Electron Energy Spectra}
\label{subsec:SIMP0136_baseline_spectra}

\begin{figure*}[ht!]
\centering
\includegraphics[width=\textwidth]{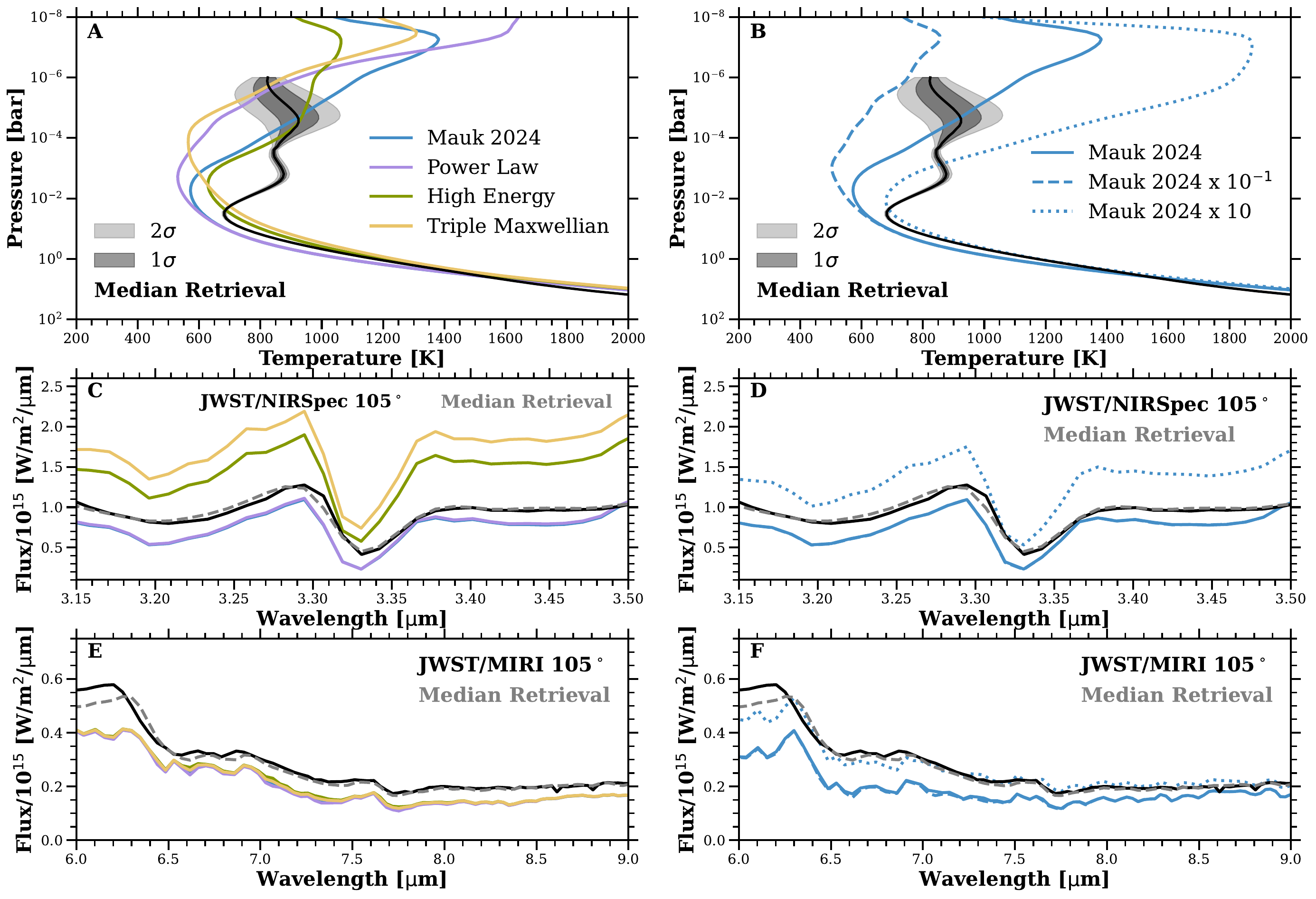}
\caption{A: Pressure–temperature profiles of SIMP0136. The median retrieved profile from \citet{Nasedkin2025} is shown in black, with the 1$\sigma$ and 2$\sigma$ uncertainty regions shaded in gray. Solid colored lines show \texttt{PICASO} models with different energy deposition profiles. 
B: Same as panel A, but for models using the \citet{Mauk2024} beam energy distribution scaled by different factors. 
C, D: JWST NIRSpec/PRISM spectrum of SIMP0136 at a phase angle of 105\textdegree\ (black) in the $\sim$3.33~$\upmu$m CH$_4$ region, compared to the best-fit retrieved spectrum with a thermal inversion (gray) and model spectra with different energy deposition profiles. 
E, F: JWST MIRI/LRS spectrum at the same phase (black), probing the 6–9~$\upmu$m region, compared to the corresponding model spectra.}
\label{fig:SIMP0136beam}
\end{figure*}

Figure \ref{fig:SIMP0136beam} shows the median retrieved thermal profile (black) for the JWST time-series observation at all phases of SIMP0136 from \citet{Nasedkin2025} along with the one and two-sigma regions in the two shades of gray. There are two apparent thermal inversions present in the thermal structure. The first near 10$^{-3}$ bar, and the second further up near 10$^{-5}$ bar. From the JWST observations, the contribution function from the retrievals show constraints on the deeper thermal inversion feature but no sensitivity to any part of the atmosphere above that pressure layer. The thermal inversion near 10$^{-5}$ bar is likely driven by the priors in the retrieval (E. Nasedkin, private communication). Therefore, for this work we focus on the thermal inversion near 10$^{-3}$ bar. Further discussions on the limitations of the retrieved profiles and our forward models are in Section \ref{sec:forward_retrievals}. We find that no electron beam energy model tested reproduces the observed temperature inversion for SIMP0136. As seen in panel B, no scaling of the \citet{Mauk2024} profile can reproduce the retrieved thermal inversion as well.

Looking at the JWST observations of the CH$_4$ features both with NIRSpec/PRISM (panel C and D) and MIRI/LRS (panel E and F), no single model is capable of matching the observations (black) or the retrieved model (gray). In all cases, we show the JWST observations at a phase angle of 105\textdegree~ since this was the observation that produced the maximum inversion strength in \citet{Nasedkin2025}. We note that even the lowest inversion strength phase angle of 210\textdegree~does not improve the fits of our energy injected models. Only the \citet{Mauk2024} profile scaled up by a factor of 10 comes close to matching any of the observations. The scaled up model recreates the observed amplitude and features in MIRI/LRS (panel F). All other models are too faint in this region. 

\subsubsection{Response to Ad-Hoc Heating Profile}
\label{subsec:SIMP0136_ad_hoc_spectra}

\begin{figure}[ht!]
\centering
\includegraphics[width=\columnwidth]{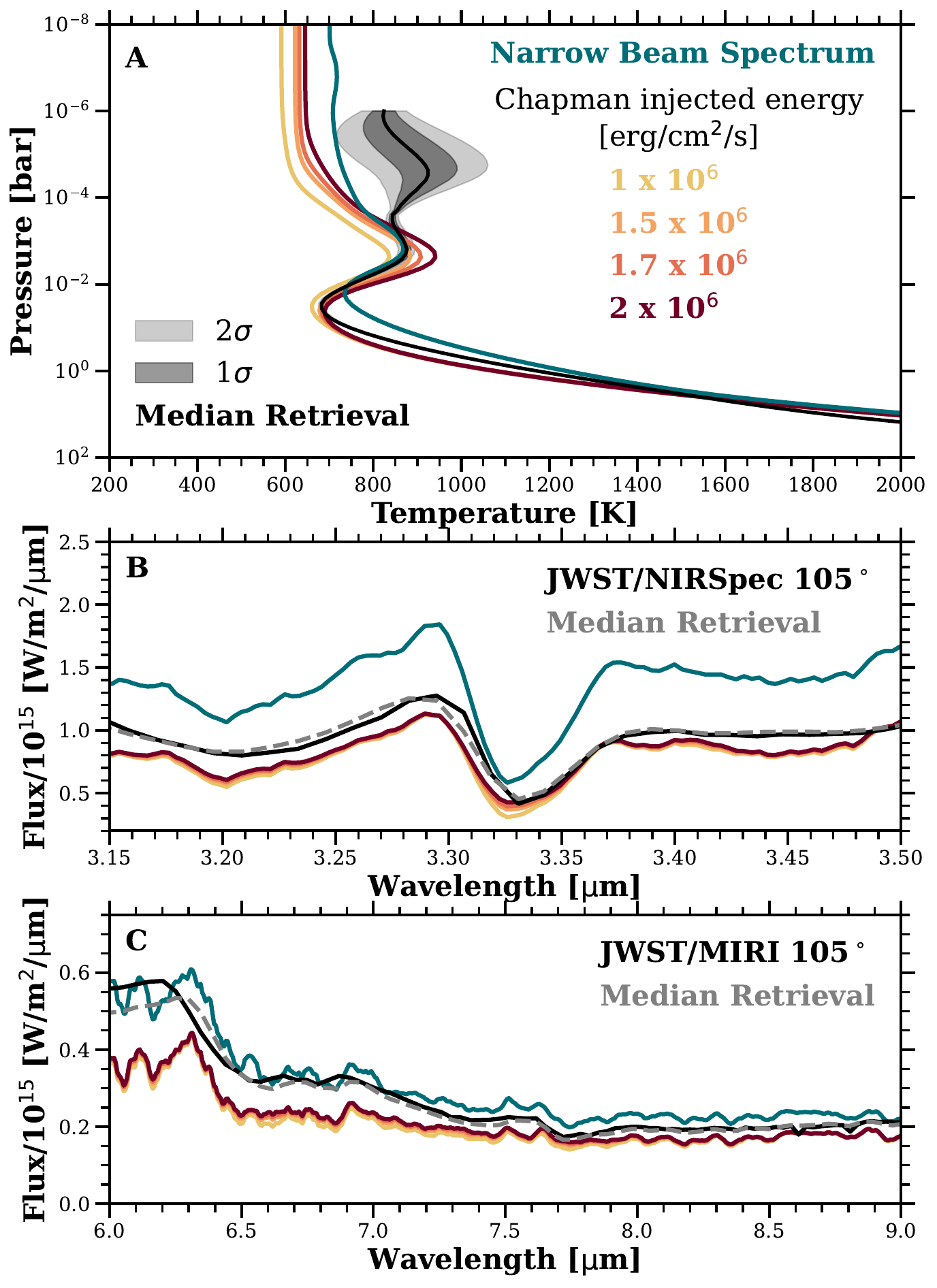}
\caption{A: Simulated pressure–temperature profiles of SIMP0136 with energy injected via a Chapman heating profile peaking at $2 \times 10^{-3}$ bar. Each profile is labeled by the total injected energy flux (erg cm$^{-2}$ s$^{-1}$). The profile with a modified, narrow electron beam spectrum is shown in blue.
B: JWST NIRSpec/PRISM spectrum of SIMP0136 at a phase angle of 105\textdegree\ (black), compared to the best-fit retrieved spectrum with a thermal inversion from \citet{Nasedkin2025} (gray) and model spectra with varying levels of energy injection. 
C: JWST MIRI/LRS spectrum at the same phase (black), probing the 6–9~$\upmu$m region, compared to the corresponding model spectra.}
\label{fig:simp0136chapman}
\end{figure}

When using the ad hoc energy deposition profile, similar to WISE1935, we are able to reproduce the required inversion at $\sim 2 \times 10^{-3}$ bar with a total energy of $1.5 \times 10^6$ erg cm$^{-2}$ s$^{-1}$ ($\sim 1\%$ of the bolometric flux of SIMP0136, Figure \ref{fig:simp0136chapman}). While this profile is able to match the thermal inversion relatively well, differences still remain in the amplitude of the wings of the CH$_4$ feature in the near-IR and at shorter wavelengths in the mid-IR.

The differing behavior between WISE1935 and SIMP0136 suggests that the response to atmospheric heating is strongly dependent on the underlying thermal structure and chemical state of the atmosphere. WISE1935, with its cooler effective temperature, has a higher intrinsic CH$_4$ abundance, allowing relatively modest heating at the appropriate pressure level to produce a pronounced emission feature. The higher effective temperature of SIMP0136 also requires more energy deposited to generate a thermal inversion in comparison to WISE1935.

\subsubsection{Beam Modification to Reproduce Required Heating}
\label{subsec:SIMP0136_mods}
As for WISE1935, we attempted to modify the baseline electron energy spectrum to reproduce the retrieved temperature profile for SIMP0136. In this case, we were also able to reproduce the inversion only with a very high energy and very narrow energy spectrum, in this case peaking at 1025keV, spanning 0.95 to 1.05 the value of the peak in range, and scaled up by an additional factor of 2.7 from the baseline spectrum. As shown in Figure \ref{fig:simp0136chapman}, this energy spectrum results in a temperature profile which matches the retrieved profile well near the maximum of the inversion, but is overall hotter at higher and lower pressures.

\subsection{Spatial Heterogeneity} \label{sec:spatial}
Our models are one-dimensional and therefore are only able to simulate the effects of energy deposition locally on the atmosphere in the region of the energy deposition. However, real atmospheres are spatially heterogeneous, and auroral energy is deposited into a polar region surrounded by a non-aurorally heated atmosphere. We can approximate this by combining the contribution of spectra from both auroral and non-auroral regions of the atmosphere. The observed spectrum is a superposition of the spectra emerging from each region of the object. Therefore, we can approximate the presence of a polar aurorally heated region and a non-heated region by constructing a linear combination of the simulated spectra from the aurorally heated atmosphere and a non-heated atmosphere.

\begin{figure*}
    \centering
    \includegraphics[width=0.48\linewidth]{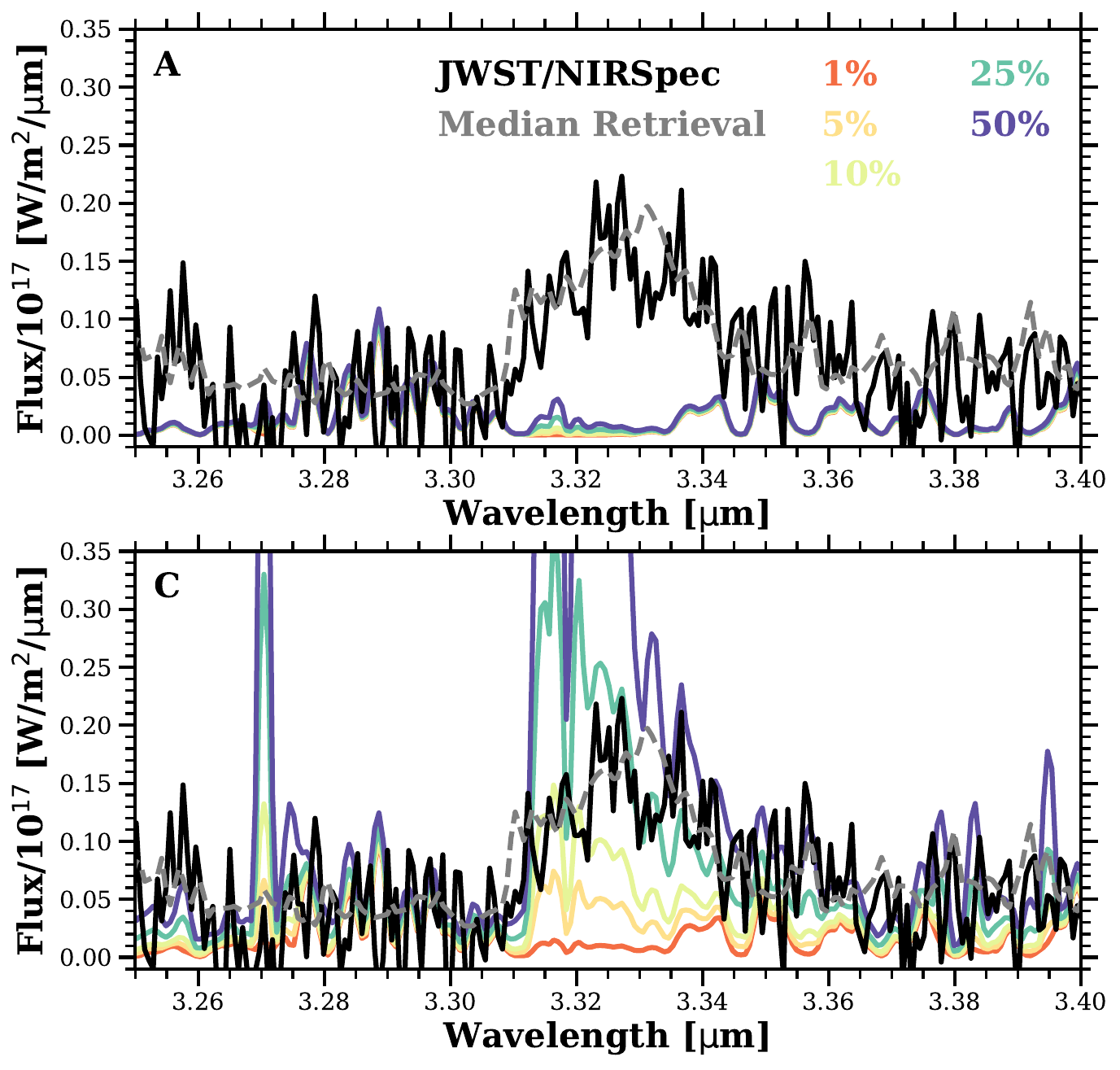}
    \includegraphics[width=0.48\linewidth]{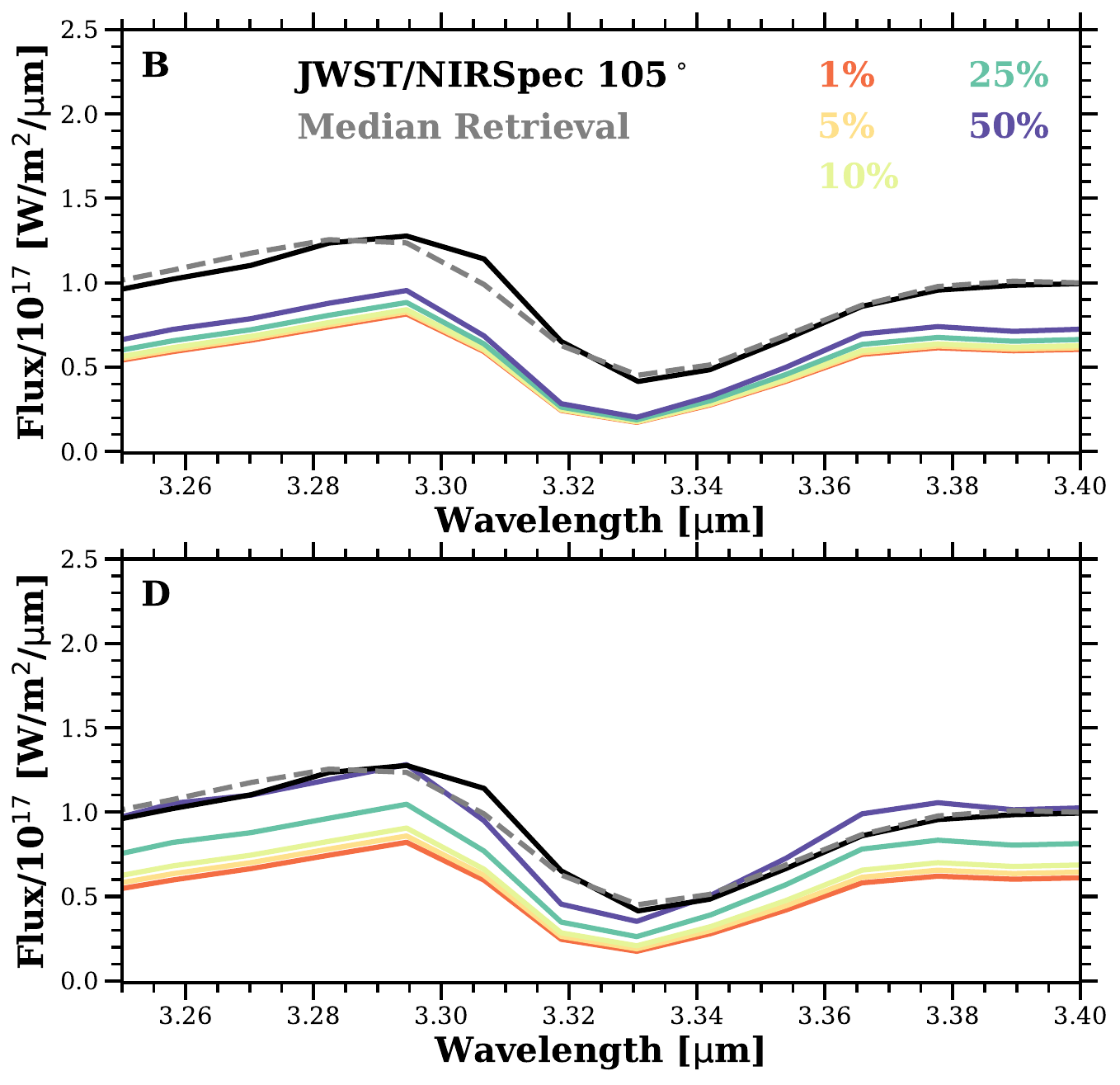}
    \caption{Spectra of WISE1935 (left) and SIMP0136 (right) in the $\sim$3.3~$\upmu$m CH$_4$ bandpass. JWST/NIRSpec observations are shown in black, while colored lines indicate models with varying fractions of auroral coverage, and the gray lines show the median retrieved spectrum. Panels A and B use the \citet{Mauk2024} energy distribution, while panels C and D show the same distribution scaled up by a factor of 10.}
    \label{fig:fraction_spectra}
\end{figure*}

Figure~\ref{fig:fraction_spectra} shows the combined spectra for varying fractions of auroral coverage for WISE1935 (panels A and C) and SIMP0136 (panels B and D). The top panels use the fiducial \citet{Mauk2024} energy deposition profile to represent the auroral atmosphere, while the bottom panels show models in which this profile is scaled up by a factor of 10. The fiducial models do not deposit sufficient energy to produce the strong thermal inversion required to reproduce the observed CH$_4$ emission features, while increasing the total deposited energy yields thermal structures that more closely match the observed emission levels.

However, spatial heterogeneity alone cannot reconcile the discrepancies between the auroral models and the observations. For WISE1935, no combination of auroral and non-auroral regions is able to reproduce the shape of the observed CH$_4$ emission feature, even when the total flux level is approximately matched with modest coverage of $\sim$10-25\%. For SIMP0136, a fractional auroral coverage of $\sim$50\% can reproduce the amplitude of the observed CH$_4$ feature.

It is also possible that latitudinal energy transport could impact the effects of energy deposition. Such transport would smooth the temperature differences between heated and non-heated portions of the atmosphere, with our fractional linear combinations providing a rough approximation to this overall impact. Because we do not model latitudinal transport, these results represent the effect of energy deposition purely into a polar region. These results motivate the need for models that can accommodate latitudinal energy transport.

\section{Discussion}
\label{sec:discussion}

\subsection{High-Energy Component of Beam Spectrum} \label{sec:highenergy}

As demonstrated in Section \ref{subsec:WISE1935}, a strong high-energy component of the electron beam would be required to deposit energy at the high pressures required by the inversion. It is possible that the Jovian auroral electron beam could extend to MeV energies \citep{Mauk2017b}, and the stronger magnetic fields and fast rotation rates of brown dwarfs could potentially produce higher-energy auroral electrons \citep{Saur2021}. Theoretical models of the magnetospheric-ionospheric coupling of auroral currents support higher-energy electron beams \citep{Nichols2012, Turnpenney2017}.  However, as shown in Sections \ref{subsec:WISE1935_mods} and \ref{subsec:SIMP0136_mods}, the presence of a higher-energy tail on its own is likely insufficient to reproduce the shape of the retrieved temperature profiles. Instead, a significant spike in the energy distribution at a high energy may be one way to achieve the necessary heating. This highlights the need for better models of the high-energy components of auroral electron beams for these objects, which might contribute to energy deposition at the required higher pressures.

In addition, because our auroral model is non-relativistic, we are currently limited in our ability to confidently model high-energy electrons. In particular, we are not aware of any currently available cross sections for the interaction of relativistic ($\gtrsim$ 500keV) electrons with H$_2$. If future work reveals that a strong high-energy component is in fact present, developing the framework to accurately model the interactions of these high-energy electrons will be a priority. Our simple extrapolations of the parameterization described by \cite{Zuckerman2026} indicate electrons with energies on the order of a few MeV should reach their maximum energy deposition rates at the approximate pressure levels where the inversions occur. Two relativistic effects mean that this can be considered a lower limit on the required electron energy: first, high-energy electrons will experience higher densities due to length contraction and thus interact at higher altitudes, and second, interaction cross sections are expected to be overall higher for relativistic electrons (L. Scarlett, private communication) again meaning interactions can happen at higher altitudes. Developing a relativistic model will be important should a significant high-energy component of the auroral electron beam be indicated in the future.

\subsection{Importance of Disequilibrium Chemistry} \label{sec:deq-chem}
Given that disequilibrium chemistry is ubiquitous in substellar atmospheres \citep{Miles2020, Zhang2025}, a self-consistent treatment of vertical mixing is required to accurately model these systems. This is supported by the recent retrieval study of SIMP0136 \citep{Nasedkin2025}, which finds that disequilibrium chemistry is necessary to reproduce the observed abundances of key species such as CH$_4$, CO, H$_2$O, H$_2$S, and PH$_3$.

The abundance of molecules such as CH$_4$ in the observable upper atmosphere plays a critical role in shaping the spectral signatures of heating. In chemical equilibrium, the relative abundance of CH$_4$ and CO is set by the local temperature, with CH$_4$ favored at cooler temperatures and CO at hotter temperatures \citep{Marley2015}. However, vertical mixing can drive the atmosphere out of equilibrium by transporting material from deeper, hotter layers upwards, resulting in quenched abundances that differ from local equilibrium predictions \citep{Marley2015, Mukherjee2022, Mukherjee2024}. Specifically in this case, disequilibrium chemistry decreases the CH$_4$ abundance in the upper atmosphere. 

This distinction is especially important in the presence of auroral energy deposition. The heating associated with auroral processes alters the local thermal structure. If equilibrium abundances are assumed, there is higher CH$_4$ opacity at low pressures, which leads to excessive absorption and suppresses the emergent flux, limiting the ability to reproduce the observed emission features. In contrast, reduced CH$_4$ abundances from disequilibrium chemistry allow the heated layers to contribute more effectively to the emergent spectrum, producing stronger emission signatures. 

The strength of vertical mixing, parameterized by $K_{\rm zz}$, directly impacts the abundances of species such as CH$_4$. We adopt a constant value of $K_{\rm zz} = 10^7$ cm$^2$ s$^{-1}$ throughout the atmosphere, which lies within the range typically inferred for substellar objects, although the vertical structure and magnitude of $K_{\rm zz}$ remain uncertain \citep{Mukherjee2022}. This choice is further supported by the retrieval study of SIMP0136, which finds comparable mixing strengths \citep{Nasedkin2025}. A detailed comparison between equilibrium and disequilibrium chemistry models will be presented in a forthcoming study.

\subsection{Origin of the CH$_4$ Emission Feature Shape} \label{sec:methane_lineshape}

When comparing the observed JWST spectrum of WISE1935 to the simulated spectra from the electron beam and ad hoc energy deposition models, clear differences emerge in the CH$_4$ band between 3.25–3.4 $\upmu$m. The first prominent distinction occurs at 3.33–3.35 $\upmu$m, where the high-energy electron beam model exhibits an extended, asymmetric tail, while the ad hoc energy deposition model produces a smoother, Gaussian-like profile that more closely matches the observed spectrum. Additionally, the electron beam models show a sharp emission spike near 3.27 $\upmu$m that is not present in the data.

\begin{figure*}
    \centering
    \includegraphics[width=\linewidth]{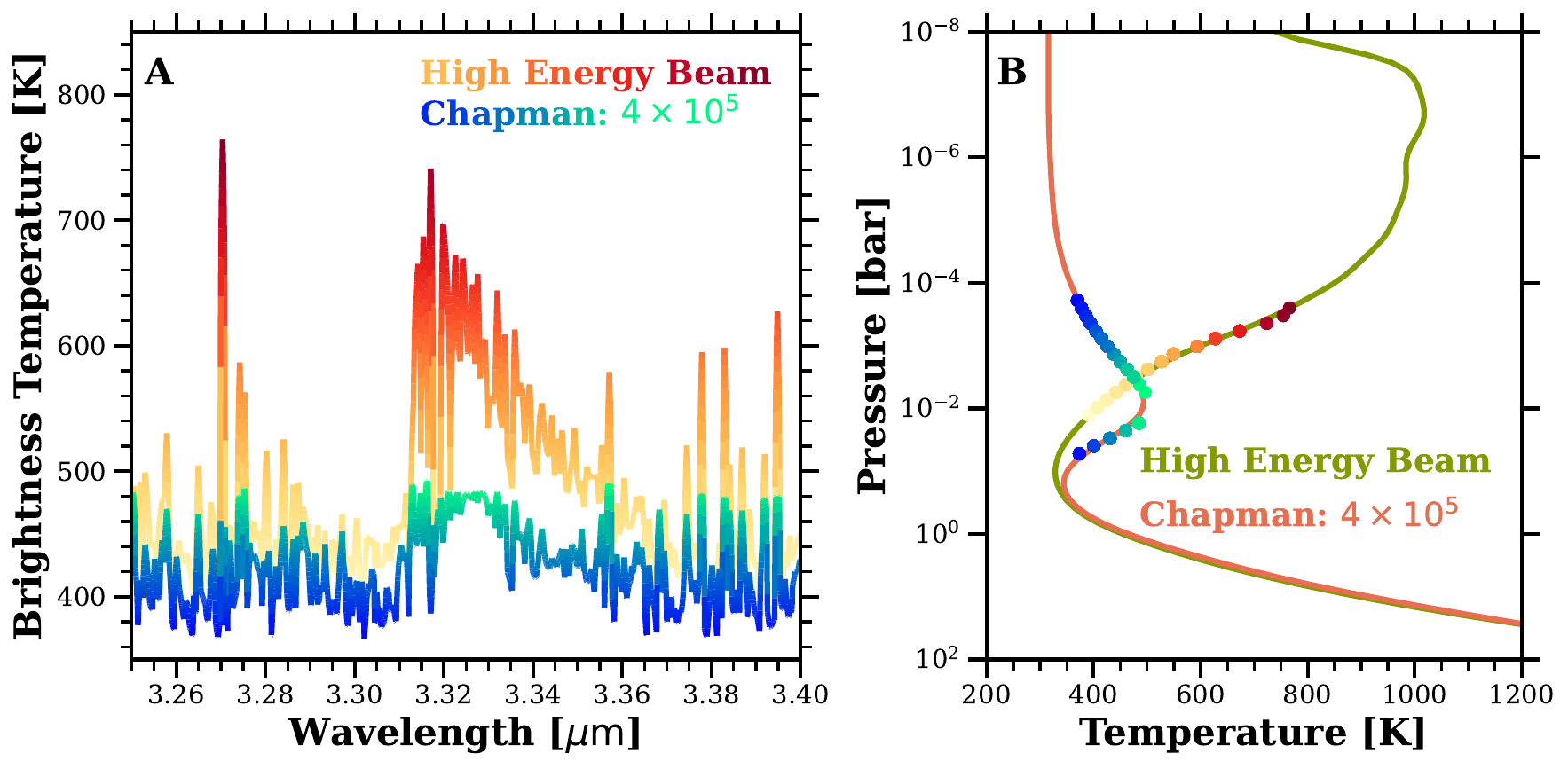}
    \caption{A: Brightness temperature spectra across the 3.25–3.4~$\upmu$m CH$_4$ band for the high-energy electron beam and Chapman function model (with 4 x 10$^5$ ergs cm$^{-2}$ s$^{-1}$ total energy injected) for WISE1935. B: Corresponding pressure–temperature profiles, with colored markers indicating the atmospheric layers probed at each wavelength. The electron beam model samples hotter, lower-pressure regions, whereas the Chapman model probes deeper layers, illustrating how the CH$_4$ line shape traces the vertical location of energy deposition.}
    \label{fig:brightness}
\end{figure*}

To understand the origin of these differences, we examine the brightness temperature spectrum and where each feature probes in the atmospheric thermal profile (Figure~\ref{fig:brightness}). Panel A shows the brightness temperature as a function of wavelength, while Panel B maps these wavelengths onto the corresponding pressure levels in each model atmosphere. In the electron beam case, the brightest portions of the spectrum, including the 3.27 $\upmu$m spike, originate from hot ($\sim$700–750 K) regions in the upper atmosphere at low pressures ($\sim$10$^{-3}$–10$^{-4}$ bar). In contrast, the ad hoc energy deposition model probes deeper layers at higher pressures ($\sim$10$^{-2}$ bar), where the peak of the thermal inversion is.

The difference reflects the temperature of the atmosphere at the layer where the atmosphere becomes optically thick. For both cases, the atmosphere becomes optically thick at the most opaque wavelengths (3.27 and 3.32 microns) at about 10$^{-4}$ bar. The atmosphere will radiate at approximately the temperature at which the atmosphere becomes optically thick. The high-energy beam model is 800 K at 10$^{-4}$ bar, creating a strong emission feature. The ad hoc energy injection model is instead about 400 K at 10$^{-4}$ bar, muting the feature. The shape of the methane feature thus allows us to pinpoint the location of the temperature inversion. 

These results indicate that the observed CH$_4$ emission on WISE 1935 is unlikely to originate from auroral heating confined to the upper atmosphere. Instead, the line shape requires energy deposition at deeper pressures, where the atmosphere becomes optically thick at these wavelengths, pointing to a heating mechanism distinct from high-altitude auroral processes. More broadly, the spectral line shape provides a diagnostic of the vertical location of thermal inversions in these atmospheres.

\subsection{Tension Between Retrievals and Forward Models} \label{sec:forward_retrievals}

Retrieval frameworks produce statistically driven models that reproduce the observed spectra with high fidelity. However, because retrieval constraints are inherently limited by the contribution functions of the observed wavelengths, only the regions of the atmosphere probed by the data are well constrained. The spectral features used to infer the thermal inversion primarily probe pressures above the deep atmosphere, and are therefore largely insensitive to the details of the thermal structure at higher pressures. This is reflected in the retrieved temperature profiles, where the 1$\sigma$ and 2$\sigma$ uncertainties are smallest in the mid-atmosphere and increase toward both lower and higher pressures due to reduced sensitivity.

Additional differences arise from the choice of parameterization within each retrieval framework. The methods used to derive the thermal structure for WISE 1935 \citep[\texttt{Brewster};][]{Burningham2017, Burningham2021} and that for SIMP0136 \citep[\texttt{petitRADTRANS};][]{Molliere2019} adopt different assumptions, most notably in the treatment of the deep atmosphere. The SIMP0136 retrieval in \citet{Nasedkin2025} enforces a fully convective deep atmosphere, whereas the WISE1935 retrieval places no such constraint. As a result, the SIMP0136 deep atmosphere ($P \gtrsim 0.1$ bar) is in closer agreement with our forward models, which assume a convective adiabat below the radiative–convective boundary as determined by the RCE solver.

In contrast, the retrieved thermal structure for WISE1935 exhibits a steeper gradient in the deep atmosphere that deviates from an adiabat, potentially resembling non-adiabatic convective profiles explored in previous studies \citep{Tremblin2015, Tremblin2019}. Despite these differences, the spectral regions analyzed here probe pressures above these deeper layers (Section~\ref{sec:methane_lineshape}), and are therefore largely insensitive to the details of the deep atmospheric structure. Consequently, the discrepancies between retrieval and forward model thermal profiles at high pressures do not impact our interpretation of the CH$_4$ emission features. Future observations at both shorter and longer wavelengths, which probe deeper and higher up into the atmosphere, respectively, will be required to further constrain these regions.

\subsection{Additional Potential Heating Mechanisms}
We demonstrate in this work that auroral heating on its own does not directly produce the inferred atmospheric inversions in the atmospheres of WISE1935 and SIMP0136, although accounting for spatially heterogeneous atmospheres in the T-dwarf may resolve some of this tension.  Several other mechanisms have been proposed to provide this heating, but further work is needed to assess whether any could contribute sufficient energy. 

One mechanism which is important for gas giants and may therefore be important for brown dwarfs is Joule heating \citep{Bougher2005}. The auroral electron beam is part of a current system which exists between the brown dwarf magnetosphere and atmosphere. The imperfect conductance of the atmosphere means that there will be resistive heating due to this current. Ionization due to auroral precipitation increases the magnitude of this Joule heating through its effect on the atmospheric conductivity. This means that even though auroral processes cannot directly produce sufficient heating for an inversion, they may be an important factor in producing additional heating. 

Another possibility is heating due to vertically traveling gravity waves \citep{Achilleos1998, Bougher2005}.  Gravity waves, or vertically propagating oscillations in the atmosphere, can transport energy from lower layers. However, this may be insufficient on its own to produce the observed heating of the Jovian thermosphere \citep{Bougher2005}. \citet{Smith2026} explored additional heating mechanisms (e.g. Joule heating and cometary impacts) and their potential contributions to producing the observed thermal inversion on WISE1935.

\subsection{Limitation of 1D Atmospheric Modeling}
\label{subsec:1Dvs3D}
As mentioned in Section \ref{subsec:atmospheric_modeling}, the \texttt{PICASO} models used in this work are 1D and therefore cannot fully resolve the multidimensional transport of energy and material through the atmosphere as a 2D or 3D model would. This limitation is particularly relevant for auroral heating because auroral energy deposition is expected to be localized at high latitudes and circulation then redistributes the heat towards other parts of the atmosphere. The Solar System objects provide useful analogs for this effect. In Earth's atmosphere, equatorial upwelling and high-latitude downwelling contribute to the global redistribution of energy \citep{Brewer1949}, and more recent General Circulation Models (GCMs) continue to demonstrate the importance of this large-scale overturning circulation \citep[e.g.][]{Butchart2014}. For Jupiter, GCMs that include auroral and Joule heating predict strong circulation away from the auroral regions, redistributing auroral heat toward the equator on timescales of $\sim 4$--$10$ Earth days at pressures below $\sim 0.15~\upmu$bar, where downwelling can contribute to the thermal balance at lower latitudes \citep{Bougher2005, Majeed2005, Majeed2009}. GCMs generated for Saturn similarly show that Joule heating can produce enhanced temperatures while vertical winds regulate how efficiently this energy is redistributed away from the auroral region \citep{Smith2005}. These Solar System examples demonstrate that higher-dimensional models are able to study the latitude-dependent thermal redistribution from polar auroral features and vertical circulation patterns that cannot be represented by a single 1D model like those presented in this work.

For brown dwarfs and giant exoplanets, studies have similarly shown that jets, turbulence, tracer transport, and cloud inhomogeneities can arise in brown dwarf atmospheres, with the resulting effective vertical mixing depending on the circulation, tracer lifetimes, particle sizes, and radiative or frictional damping \citep{TanShowman2021a,TanShowman2021b}. This can result in horizontal temperature variations of $\sim 50~$K \citep{Showman2013}.

We have attempted to approximate the disk-integrated spectral impact of this spatial inhomogeneity by combining heated and unheated spectra, as discussed in Section \ref{sec:spatial}. However, this approach does not capture the circulation-driven redistribution of deposited energy. Future work using 2D models with localized polar heating, and ultimately 3D time-dependent models coupled to radiative transfer and chemistry, will help to determine whether auroral circulation can transport energy to the pressures required by the observed CH$_4$ emission.

\section{Conclusions} \label{sec:conclusion}
In this work we investigated whether auroral electron precipitation can account for the thermal inversions inferred in the atmospheres of WISE1935 and SIMP1036 from JWST observations and retrievals. By coupling physically motivated auroral energy deposition profiles to radiative-convective equilibrium models, we find:

\begin{enumerate}[label=(\arabic*)]
    \item Jovian-like auroral electron precipitation is unable to reproduce the observed thermal inversions in either object. While auroral heating generates inversions, the energy is deposited at pressures that are too low compared to those required by the retrievals.
    \item Increasing the total energy flux of the electron beam strengthens the inversion but does not shift it to deeper pressures, demonstrating that the vertical location of heating is primarily controlled by the electron energy distribution rather than the total deposited energy. Improved models, beyond the simple extrapolations presented here, which can accurately characterize the relativistic behavior of high-energy electrons will be critical for assessing the impact of these electrons on atmospheric heating.
    \item Parameterized ad hoc heating profiles (Chapman functions) can reproduce both the amplitude and pressure location of the thermal inversions but require energy deposition at much greater depths in the atmosphere than those produced by the sample of auroral models. This suggests that a high-energy beam component and/or alternative heating mechanisms are required to reproduce the retrieved thermal inversions. 
    \item The shape of the CH$_4$ emission feature provides a direct diagnostic of the vertical location of energy deposition. Models in which heating occurs at low pressures produce sharp line shapes that are not present in the data, whereas deeper heating produces the broader spectral profile observed in WISE1935.
    \item The differences in our ability to reproduce the thermal structure and spectral features of SIMP0136 compared with WISE1935 indicate that additional physical processes, or differences in atmospheric structure between the two objects, must play an important role.
    \item Integrated modeling efforts will be needed to fully characterize the consequences of auroral precipitation on brown dwarf atmospheres. Though auroral heating on its own does not reproduce the inversions, auroral precipitation has additional impacts which could provide avenues for further heating. In particular, auroral ionizations could drive Joule heating. This highlights the need for the integration of a broad range of physical models to fully assess indirect auroral impacts on the heating of brown dwarf atmospheres.
\end{enumerate}

These results rule out auroral heating, confined to the upper atmosphere, as the sole origin of the temperature inversions for WISE1935 and SIMP0136. Instead, the retrieved thermal inversion requires energy deposition at deeper pressures, pointing to a heating mechanism distinct from direct auroral processes. Future work incorporating relativistic particle populations and additional heating mechanisms such as Joule heating or wave-driven energy transport will be essential for fully understanding the origin of these inversions.

\begin{acknowledgments}
We thank Jacqueline K. Faherty and Ben Burningham for providing their retrieval results on WISE1935 and Evert Nasedkin for helpful discussions about SIMP0136. J.M. acknowledges support from the National Science Foundation Graduate Research Fellowship Program under Grant No. DGE 2137420. A.Z acknowledges support from the National Science Foundation Graduate Research Fellowship Program under Grant No. DGE 2040434. This work was partially funded by the Space Telescope
Science Institute under awards JWST 1874 and 6474.
Any opinions, findings, and conclusions or recommendations expressed in this material are those of the authors and do not necessarily reflect the views of the National Science Foundation.
\end{acknowledgments}

\begin{contribution}

A.Z developed the auroral models and contributed the energy flux rates resulting from this modeling, and contributed to the writing of the manuscript.  
J.M. developed the \texttt{PICASO} functionality to inject additional energy into the atmosphere, generated all the atmospheric models and synthetic spectra, and contributed to the writing of the manuscript.
J.S.P and C.M helped conceive the project, guided direction, and interpretation of results, while contributing to the writing. 
D.B contributed valuable initial discussion and provided feedback on the manuscript. 
\end{contribution}

\software{\texttt{PICASO} \citep{Batalha2019,Mukherjee2023,Mang2026}, \texttt{Jupyter} \citep{kluyver2016jupyter}, \texttt{NumPy} \citep{walt2011numpy}, \texttt{SciPy} \citep{SciPy}, and \texttt{Matplotlib} \citep{matplotlib}.}

\bibliography{works_cited}
\bibliographystyle{aasjournal}

\appendix
\FloatBarrier

\section{High-Energy Extrapolation}
\label{sec:extrapolation}
\begin{figure*}[ht!]
    \centering
    \includegraphics[width=\linewidth]{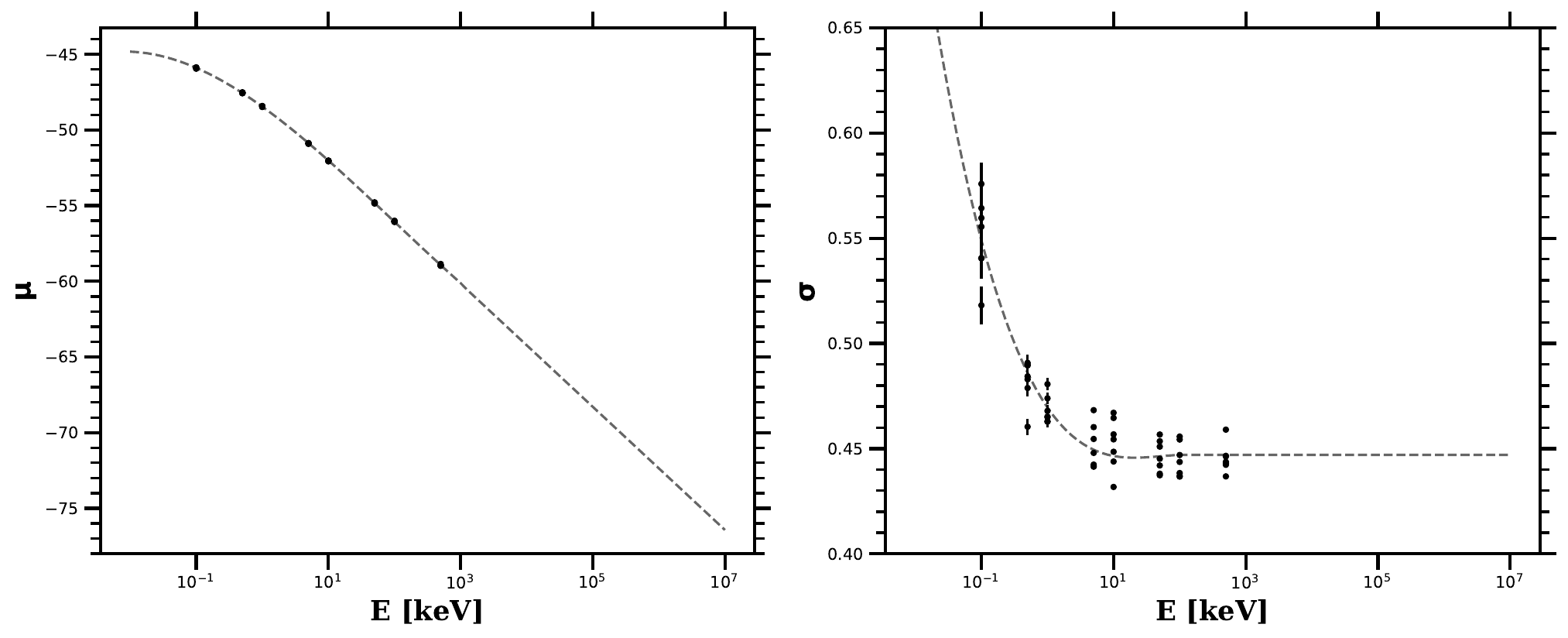}
    \caption{Extrapolations (dashed) for the energy-dependent parameters of the \cite{Zuckerman2026} heating profile parameterization. Black points represent fit values from that paper with one-standard deviation errorbars shown.}
    \label{fig:mu_and_sigma_extended}
\end{figure*}

In Sections \ref{subsec:WISE1935_mods} and \ref{subsec:SIMP0136_mods} we have approximated the behavior of electron beams with very high energy components. As discussed in Section \ref{sec:highenergy}, our model does not accurately treat these relativistic electrons. We have simply extrapolated the parameterization (Eq. \ref{eq:qenergy}) of the heating rate $q_{\varepsilon}$ derived from non-relativistic electron energies to higher energies. The details of the parameterization are discussed in \cite{Zuckerman2026}, and the energy dependence is parameterized through the terms $\mu(\varepsilon_0)$ and $\sigma(\varepsilon_0)$ (Equation 3 in that paper). In that paper, these parameters were fit from the results of simulations spanning 0.1-500keV, and the functions used do not extrapolate well to higher energies due to their polynomial forms. In this work we have extended these in a piecewise fashion above 500keV as shown in Figure \ref{fig:mu_and_sigma_extended}. We fit a linear extension for $\mu(\varepsilon_0)$.  $\sigma(\varepsilon_0)$, which approaches a constant value at high energies, we simply take to be constant above 500keV.

\section{Additional Heating Profiles}
\label{sec:modified_beam_heating}
\begin{figure*}[ht!]
    \centering
    \includegraphics[width=0.5\columnwidth]{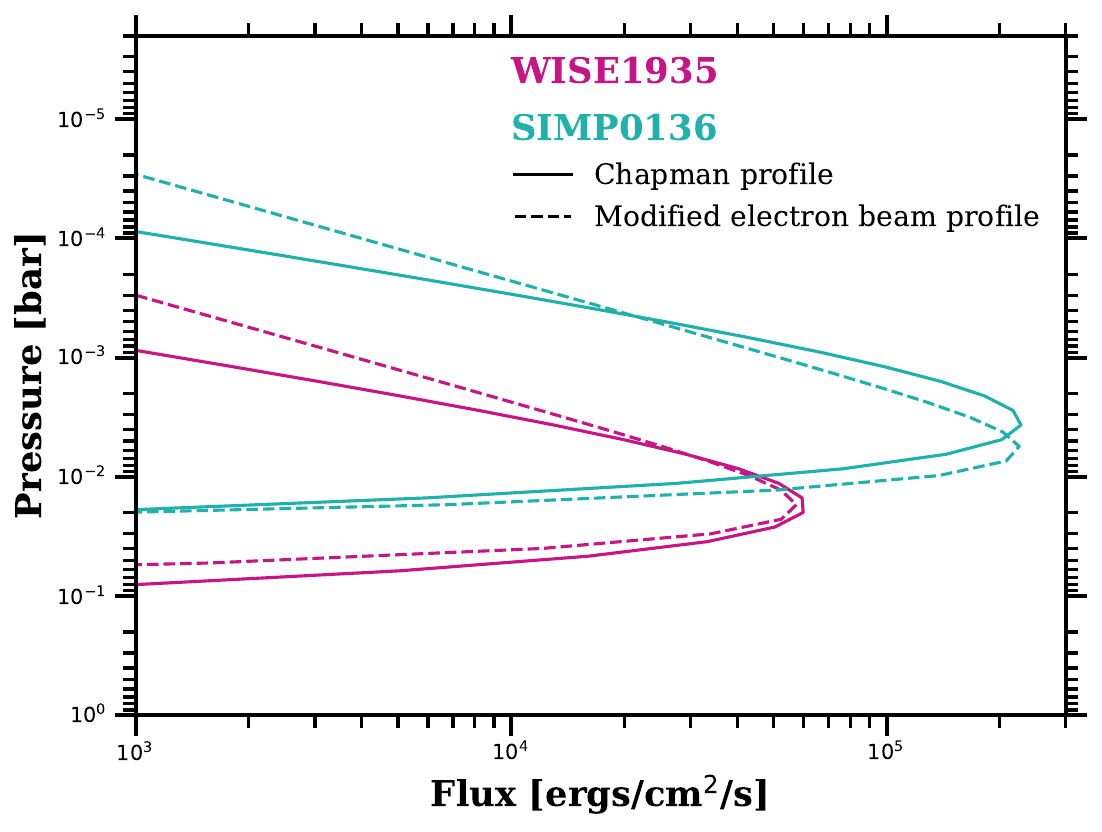}
    \caption{Heat flux resulting from the best-fitting ad hoc Chapman profile compared with heat flux resulting from the modified narrow beam spectrum which best reproduces the location of the inversion in the PT profile for each object. The best fitting Chapman profile for WISE1935 deposits about $4 \times 10^{5}$ erg~cm$^{-2}$~s$^{-1}$, and the best fitting Chapman profile for SIMP0136 deposits about $1.5 \times 10^6$ erg cm$^{-2}$ s$^{-1}$. }
    \label{fig:chapman_and_modified_beam_heat_flux}
\end{figure*}

In Sections \ref{subsec:WISE1935_mods} and \ref{subsec:SIMP0136_mods} we attempt to modify the electron beam energy spectrum to produce the required heating profiles. The heat flux profiles resulting from the electron beams that best reproduce the inversions in the temperature profiles are shown in Figure \ref{fig:chapman_and_modified_beam_heat_flux}. For WISE1935, the beam which best reproduces the required Chapman heat flux profile also best reproduces the location of the inversion in the temperature profile. However, for SIMP0136 the beam spectrum that best reproduces the inversion must produce a peak in heat flux at a somewhat higher pressure than does the required Chapman profile. Heating occurs at lower pressures for SIMP0136, in the radiative zone where it is difficult to dissipate the heat effectively. Additionally, more energy overall is required for SIMP0136, meaning more energy is deposited in the higher altitude tail where the modified electron beam heat profile diverges from the Chapman heat profile. Low-pressure layers are more responsive to the input of heat. This means that the more efficient temperature increase of overlying low-pressure layers is sufficient to shift the location of maximum temperature increase upwards relative to the location of maximum heat input for SIMP0136. Because the heating for WISE1935 occurs in a higher-pressure region where radiative cooling is still effective, this impact is less significant.

\end{document}